\documentclass[aps,prl,reprint,superscriptaddress,showpacs]{revtex4-1}

\usepackage{graphicx}
\usepackage{amsmath}
\begin{document}

% Use the \preprint command to place your local institutional report
% number in the upper righthand corner of the title page in preprint mode.
% Multiple \preprint commands are allowed.
% Use the 'preprintnumbers' class option to override journal defaults
% to display numbers if necessary
%\preprint{}

%Title of paper
\title{Geometric spin Hall effect of light in tightly focused polarization tailored light beams}

% repeat the \author .. \affiliation  etc. as needed
% \email, \thanks, \homepage, \altaffiliation all apply to the current
% author. Explanatory text should go in the []'s, actual e-mail
% address or url should go in the {}'s for \email and \homepage.
% Please use the appropriate macro foreach each type of information

% \affiliation command applies to all authors since the last
% \affiliation command. The \affiliation command should follow the
% other information
% \affiliation can be followed by \email, \homepage, \thanks as well.
\author{Martin Neugebauer}
\email[]{martin.neugebauer@mpl.mpg.de}
\homepage[]{http://www.mpl.mpg.de/}
\affiliation{Max Planck Institute for the Science of Light, Guenther-Scharowsky-Str. 1, D-91058 Erlangen, Germany}
\affiliation{Institute of Optics, Information and Photonics, University Erlangen-Nuremberg, Staudtstr. 7/B2, D-91058 Erlangen, Germany}
\author{Peter Banzer}
\affiliation{Max Planck Institute for the Science of Light, Guenther-Scharowsky-Str. 1, D-91058 Erlangen, Germany}
\affiliation{Institute of Optics, Information and Photonics, University Erlangen-Nuremberg, Staudtstr. 7/B2, D-91058 Erlangen, Germany}
\author{Thomas Bauer}
\affiliation{Max Planck Institute for the Science of Light, Guenther-Scharowsky-Str. 1, D-91058 Erlangen, Germany}
\affiliation{Institute of Optics, Information and Photonics, University Erlangen-Nuremberg, Staudtstr. 7/B2, D-91058 Erlangen, Germany}
\author{\mbox{Sergej Orlov}}
\affiliation{Max Planck Institute for the Science of Light, Guenther-Scharowsky-Str. 1, D-91058 Erlangen, Germany}
\affiliation{Institute of Optics, Information and Photonics, University Erlangen-Nuremberg, Staudtstr. 7/B2, D-91058 Erlangen, Germany}
\author{\mbox{Norbert Lindlein}}
\affiliation{Institute of Optics, Information and Photonics, University Erlangen-Nuremberg, Staudtstr. 7/B2, D-91058 Erlangen, Germany}
\author{Andrea Aiello}
\affiliation{Max Planck Institute for the Science of Light, Guenther-Scharowsky-Str. 1, D-91058 Erlangen, Germany}
\affiliation{Institute of Optics, Information and Photonics, University Erlangen-Nuremberg, Staudtstr. 7/B2, D-91058 Erlangen, Germany}
\author{Gerd Leuchs}
\affiliation{Max Planck Institute for the Science of Light, Guenther-Scharowsky-Str. 1, D-91058 Erlangen, Germany}
\affiliation{Institute of Optics, Information and Photonics, University Erlangen-Nuremberg, Staudtstr. 7/B2, D-91058 Erlangen, Germany}

%\thanks{}
%\altaffiliation{}

%Collaboration name if desired (requires use of superscriptaddress
%option in \documentclass). \noaffiliation is required (may also be
%used with the \author command).
%\collaboration can be followed by \email, \homepage, \thanks as well.
%\collaboration{}
%\noaffiliation

\date{\today}

\begin{abstract}
Recently, it was shown that a non-zero transverse angular momentum manifests itself in a polarization dependent intensity shift of the barycenter of a paraxial light beam [A. Aiello \textit{et al}., Phys. Rev. Lett. 103, 100401 (2009)]. The underlying effect is phenomenologically similar to the spin Hall effect of light, but does not depend on the specific light-matter interaction and can be interpreted as a purely geometric effect. Thus, it was named the geometric spin Hall effect of light. Here, we experimentally investigate the appearance of this effect in tightly focused vector-beams. We use an experimental nano-probing technique in combination with a reconstruction algorithm to verify the relative shifts of the components of the electric energy density in the focal plane, which are linked to the intensity shift. By that, we experimentally demonstrate the geometric spin Hall effect of light in a focused light beam.
\end{abstract}

% insert suggested PACS numbers in braces on next line
% insert suggested keywords - APS authors don't need to do this
%\keywords{}

%\maketitle must follow title, authors, abstract, \pacs, and \keywords
\maketitle

\begin{figure}[htbp]
\includegraphics[width=0.35\textwidth]{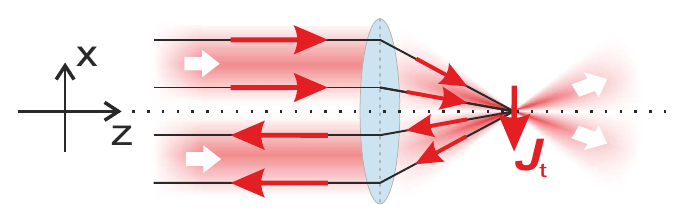}
\caption{\label{concept} Schematic illustration of the generation of light with purely transverse angular momentum. In front of the focusing system the paraxial beam propagating in positive \mbox{$z$-direction} is spatially separated in a left- and right-handed circularly polarized part along the $x$-axis. The spin (red arrows) in the upper part of the beam is parallel to the optical axis while it is anti-parallel in the lower part. Focusing of such a spin-segmented beam profile leads to the tilting of the spin vectors and, for symmetry reasons, to the generation of purely transverse angular momentum $\textbf{J}_{t}$ in the focal plane.}
\end{figure}

\begin{figure}[h!]
\includegraphics[width=0.45\textwidth]{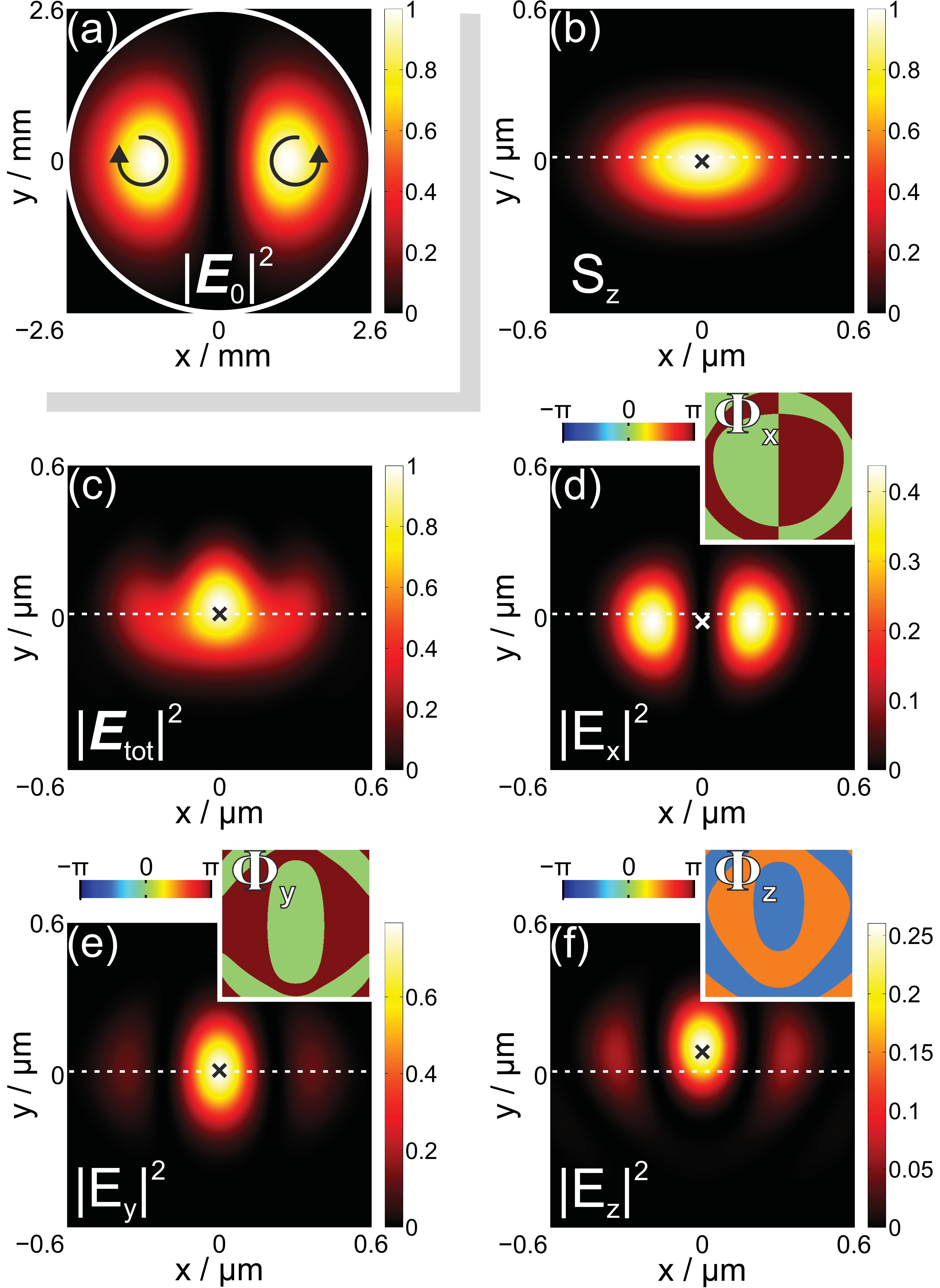}
\caption{\label{fields} (a) Normalized electric energy density distri\-bution $\left|\textbf{E}_{0}\right|^{2}$ of the incoming paraxial beam (\mbox{$\text{TEM}_{10}$-mode}, \mbox{$w_{0}\!=\!1.9$ mm}). The white ring corresponds to the geometric aperture of the microscope objective ($\text{NA}\!\!=\!1.3$). The black arrows indicate spin-segmentation. (b) shows the focal distribution of the intensity $S_{z}$, (c) the focal distribution of the electric energy density $\left|\textbf{E}_{tot}\right|^{2}$. Both distributions are normalized to their respective maximum value. (d-f) represent the components of the electric energy density $\left|E_{x}\right|^{2}$, $\left|E_{y}\right|^{2}$, and $\left|E_{z}\right|^{2}$ (normalized to the maximum value of $|\textbf{E}_{tot}|^{2}$ in (c)), and their corresponding relative \mbox{phases $\Phi_{x}$, $\Phi_{y}$, and $\Phi_{z}$} (see insets). Black and white crosses mark the position of the barycenter for each individual quantity and the dashed white line indicates the $y$-coordinate of the barycenter of $\left|\textbf{E}_{tot}\right|^{2}$ in (c) and is used as reference in (b-f).}
\end{figure}

\begin{figure}[htbp]
\includegraphics[width=0.45\textwidth]{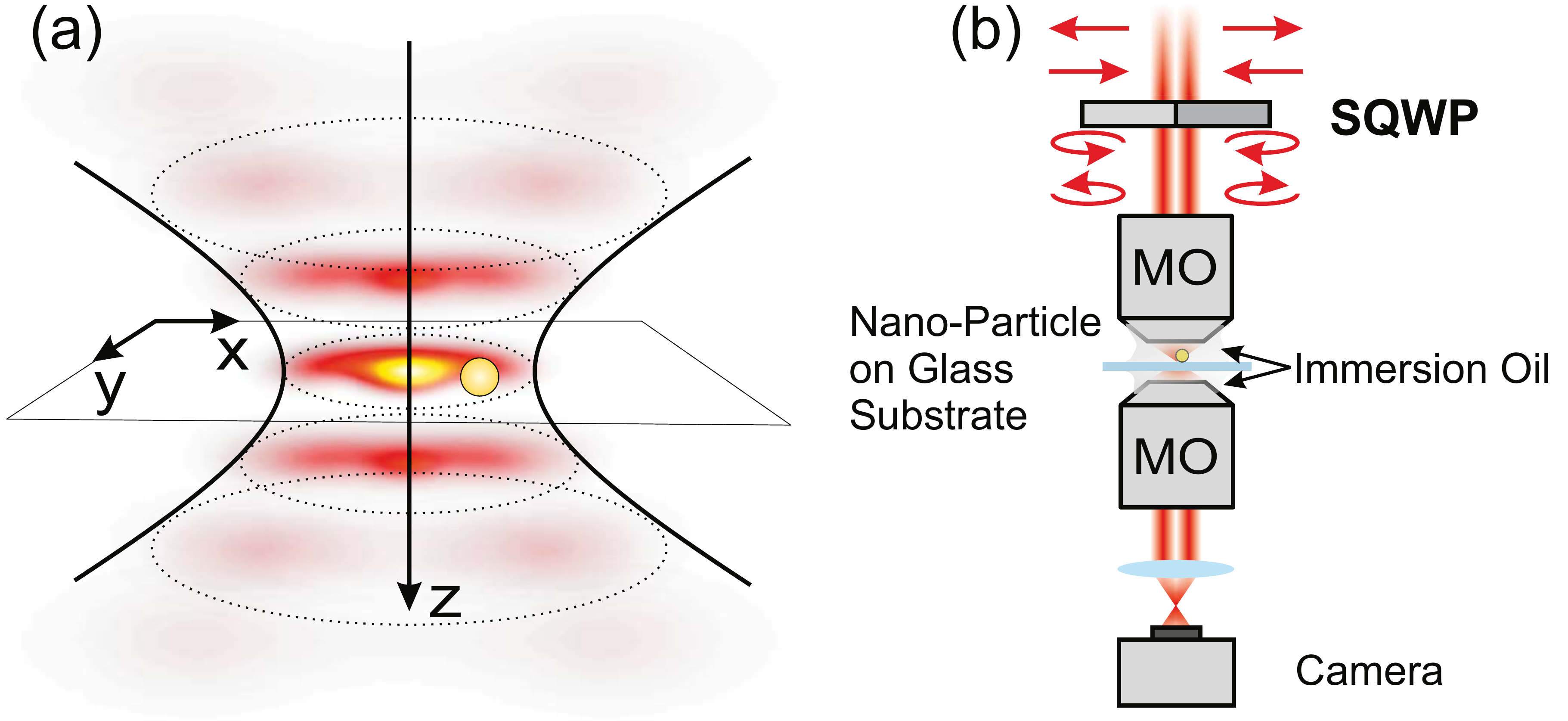}
\caption{\label{Setup} (a) Schematics of the nano-probe scanning technique. A sub-wavelength gold nano-sphere (diameter $=90$ nm) is scanned through the focal plane of the tightly focused vector-beam. At each position, the particle is excited by the local electric field. By measuring the scattered and transmitted light, the focal spot can be reconstructed. (b) Scheme of the experimental setup. The paraxial beam \mbox{($x$-polarized $\text{TEM}_{10}$-mode)} impinges on a segmented quarter wave-plate (SQWP) with perpendicular fast axes. As a result, a laterally separated left- and right-handed circularly polarized beam is generated (red arrows). This beam is coupled into a high numerical aperture immersion-type objective (MO) and focused onto the sample, consisting of a particle sitting on a glass substrate. The sample is embedded in immersion oil index matched to the substrate. A second MO, identical to the first one and with the same focal plane, is mounted below the sample. It collects the forward scattered and transmitted light. The back focal plane of the second MO is imaged with a camera.}
\end{figure}

\begin{figure}[htbp]
\includegraphics[width=0.45\textwidth]{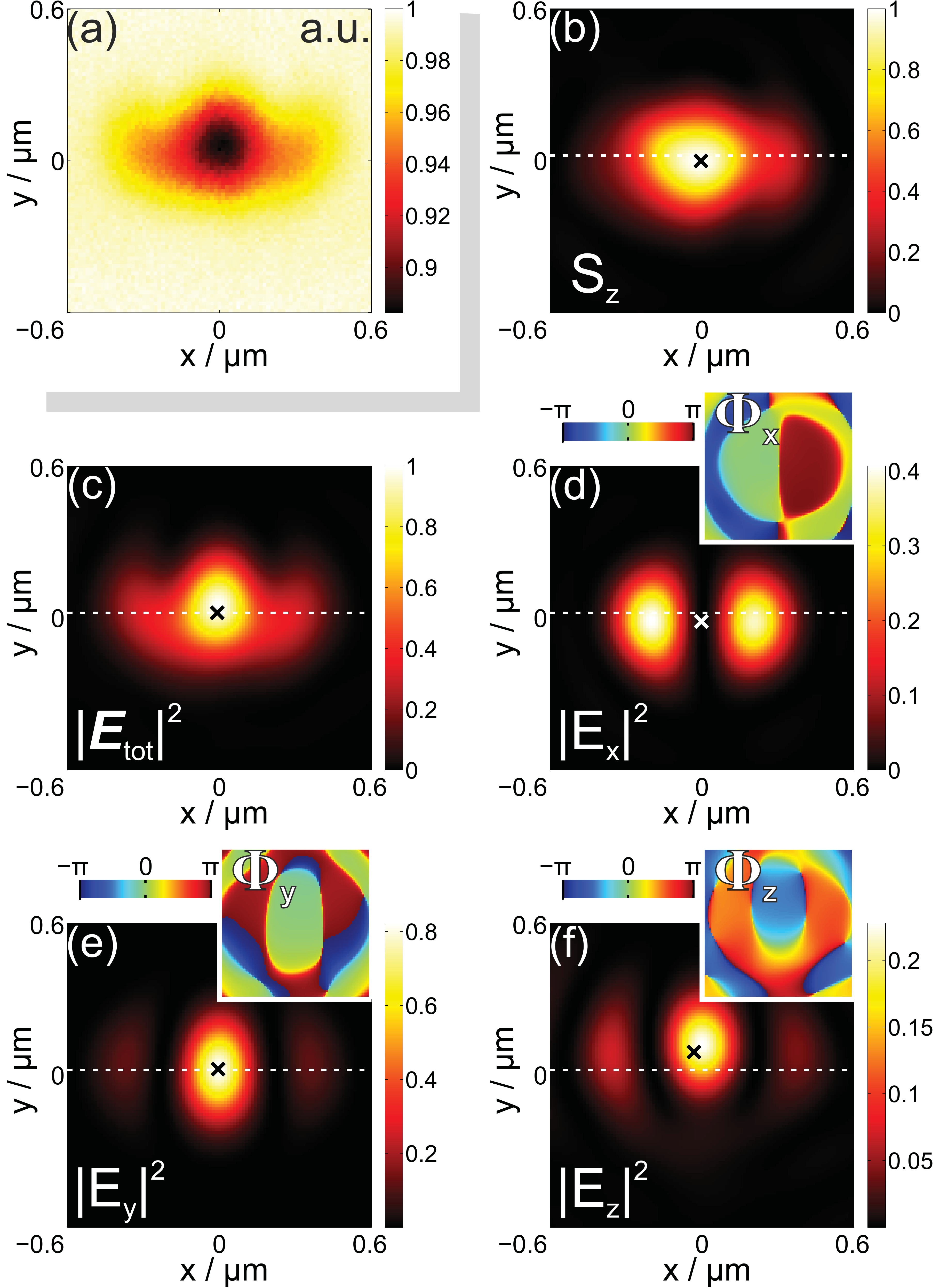}
\caption{\label{exp:results} (a) Experimental scanning result, normalized to its maximum value, measured in transmission. The signal is dominated by the extinction of the transmitted beam. In (b-f) the reconstructed distributions of $S_{z}$, $\left|\textbf{E}_{tot}\right|^{2}$, $\left|E_{x}\right|^{2}$, $\left|E_{y}\right|^{2}$, and $\left|E_{z}\right|^{2}$ are depicted similar to Fig. \ref{fields}(b-f).}
\end{figure}

% body of paper here - Use proper section commands
% References should be done using the \cite, \ref, and \label commands
%\section{Introduction one}
% Put \label in argument of \section for cross-referencing
%\section{\label{}}
%\section{\label{intro}Introduction}
\textit{Introduction.}\textemdash Angular momentum (AM) carried by a beam of light often plays a fundamental role in light-matter interaction. An important example is the Imbert-Fedorov shift, which is a manifestation of the so-called spin Hall effect of light \mbox{(SHEL) \cite{Onoda2004,Bliokh2006,Aiello2008}}. In the most basic case, the effect occurs at a planar interface between two media with different refractive indices. It describes the polarization dependent transverse spatial and/or angular shift of the propagation axis of the reflected and refracted light beam. While the SHEL is present both in reflection and in refraction, the spatial shift has first been predicted and experimentally demonstrated in total internal reflection \mbox{configuration \cite{Fedorov1955,Imbert1972}}. In the past decade, the interest concerning the SHEL has been rising. On the one hand, advancements in modern metrology enable the measurement of very small lateral displacements, which allows for utilizing the SHEL for sensing material \mbox{properties \cite{Hosten2008,Pillon2004,Jin2012,Zhou2012}}. On the other hand, the control of light in the sub-wavelength regime is crucial, for instance, in \mbox{nano-optics \cite{Rodriguez-Herrera2010,Luo2011}}.\\
Just recently, a novel type of beam shift was described, which is in many ways similar to the aforementioned phenomena, but does not depend on light-matter \mbox{interaction \cite{Aiello2009}}. This so-called geometric spin Hall effect of \mbox{light (gSHEL)} refers to a transverse shift of the barycenter of the beam intensity\textemdash defined as the \mbox{$z$-component} of the time averaged Poynting vector $S_{z}$\textemdash while the actual propagation axis of the beam\textemdash defined by the barycenter of the total energy density and the propagation direction\textemdash remains \mbox{unaffected \cite{Aiello2009}}. This shift of the barycenter of $S_{z}$ is connected to the relative shifts of the components of the electric energy \mbox{density \cite{Aiello2009,Korger2011,Korger2013}}. The gSHEL occurs when a non-zero transverse angular momentum is present, i.e. an AM component parallel to a plane of observation. Until now, this effect has only been investigated for collimated paraxial \mbox{beams \cite{Aiello2009,Korger2011,Korger2013,Kong2012}}. For the first time, we now demonstrate the appearance of the gSHEL in tightly focused vector-beams both theoretically and experimentally. We generate the necessary transverse AM by breaking the symmetry of the spin distribution of the light beam (similar to \cite{B.Ya.Zel'dovichN.D.Kundikova}). This transverse AM is actually linked to the shift of the barycenter of $S_{z}$ and the relative shifts of the components of the electric energy density. In the presented scheme, the latter causes a deformation of the focal spot, i.e. the distribution of the total electric energy density, which can be sensed by applying an appropriate nano-probe scanning technique. Furthermore, we use a recently developed amplitude and phase reconstruction algorithm \cite{Bauer2013}, to verify the link between the deformation of the focal spot and the relative shifts of the components of the energy density.\\
%%%%%%%%%%%%%%%%%%%%%%%%%%%%%%%%%%%%%%%%%%%%%%%%%%%%%%%%%%%%%%%%%%%%%%%%%%%%%%%%%%%%%%%%%%%%%%%%%%%%%%%%%%%%%%%%%%%%%%%%%%%%%%%%%%
\textit{Generation of transverse angular momentum by tight focusing.}\textemdash The basic concept of the experiment is the generation of transverse AM via focusing with a high numerical aperture (NA) microscope objective \mbox{(see Fig. \ref{concept})}.  
%%%%%%%%%%%%%%%%%%%%%%%%%%%%%%%%%%%%%%%%%%%%%%%%%%%%%%%%%%%%%%%%%%%%%%%%%%%%%%%%%%%%%%%%%%%%%%%%%%%%%%%%%%%%%%%%%%%%%%%%%%%%%%%%%%
%%%%%%%%%%%%%%%%%%%%%%%%%%%%%%%%%%%%%%%%%%%%%%%%%%%%%%%%%%%%%%%%%%%%%%%%%%%%%%%%%%%%%%%%%%%%%%%%%%%%%%%%%%%%%%%%%%%%%%%%%%%%%%%%%%
The principle can be explained within the framework of ray optics. We start by preparing a collimated monochromatic beam of light with spatially separated left- and right-handed circular polarization components \cite{Banzer2013}. The intensity profile of this spin-segmented beam is symmetric with respect to the $y$-axis and the propagation axis is the optical axis of the focusing system ($z$-axis). Such conditions are fulfilled for a beam profile described by
%%%%%%%%%%%%%%%%%%%%%%%%%%%%%%%%%%%%%%%%%%%%%%%%%%%%%%%%%%%%%%%%%%%%%%%%%%%%%%%%%%%%%%%%%%%%%%%%%%%%%%%%%%%%%%%%%%%%%%%%%%%%%%%%%%
\begin{align}
 \begin{pmatrix} 
  E_{x} \\  E_{y}
 \end{pmatrix} \propto
  \begin{pmatrix} 
   -\mathrm{i} \\  \text{sgn}(x)
 \end{pmatrix} x\,e^{-\frac{x^{2}+y^{2}}{w_{0}^{2}}}\text{,}
\label{equ:beam}
\end{align}
%%%%%%%%%%%%%%%%%%%%%%%%%%%%%%%%%%%%%%%%%%%%%%%%%%%%%%%%%%%%%%%%%%%%%%%%%%%%%%%%%%%%%%%%%%%%%%%%%%%%%%%%%%%%%%%%%%%%%%%%%%%%%%%%%%
with the polarization vector $\left(-\mathrm{i},\text{sgn}(x)\right)$ accounting for the laterally separated left- and right-handed circular polarization. The intensity profile of the beam in the entrance aperture of the microscope objective is represented by a $\text{TEM}_{10}$-mode with the width $w_{0}$. Since the incoming beam is collimated, we can neglect the $z$-component of the electric field in this representation. When the paraxial beam impinges on the focusing system, the partial rays of the beam and, consequently, the spins are redirected and therefore tilted towards the optical axis, crossing at the geometrical focus \mbox{(see red arrows in Fig. \ref{concept})}. In the focal plane, the longitudinal components of the AM cancel due to the chosen symmetry of the input distribution, whereupon the transverse components of the AM add up. Therefore, a state of light with even purely transverse AM is \mbox{generated \cite{Banzer2013}}. Thus, by defining the focal plane as the plane of observation, we expect the barycenter of $S_{z}$ to be shifted relative to the barycenter of the electric energy density of the beam. Furthermore, the indi\-vidual components of the electric energy density should be shifted relative to each other \cite{Aiello2009,Korger2011}. \\
In order to test these expectations quantitatively, we evaluate the field distribution in the focal plane for an incoming beam defined by \mbox{equation (\ref{equ:beam})}. The profile of this paraxial input beam is depicted in \mbox{Fig. \ref{fields}(a)}. For the numerical calculation of the focal fields we apply the Richards-Wolf-integrals \cite{Richards1959} and use the same optical parameters as in the experiment shown later. The wavelength $\lambda_{0}$ of the incoming paraxial beam is $590$ nm and its width $w_{0}$ is $1.9$ mm. For focusing, an aplanatic immersion-type microscope objective with an NA of $1.3$ and effective focal length $f=3$ mm is used. The medium behind the objective (immersion oil and glass) has a refractive index of $n = 1.5$.
%%%%%%%%%%%%%%%%%%%%%%%%%%%%%%%%%%%%%%%%%%%%%%%%%%%%%%%%%%%%%%%%%%%%%%%%%%%%%%%%%%%%%%%%%%%%%%%%%%%%%%%%%%%%%%%%%%%%%%%%%%%%%%%%%%
\\
%%%%%%%%%%%%%%%%%%%%%%%%%%%%%%%%%%%%%%%%%%%%%%%%%%%%%%%%%%%%%%%%%%%%%%%%%%%%%%%%%%%%%%%%%%%%%%%%%%%%%%%%%%%%%%%%%%%%%%%%%%%%%%%%%%
As already mentioned, the barycenter of the energy density lies per definition on the propagation axis of the beam. The energy density includes both, the total energy density of the electric field $\left|\textbf{E}_{tot}\right|^{2}$ and the magnetic field $\left|\textbf{B}_{tot}\right|^{2}$\cite{Berry2009}. However, here we restrict ourselves to the electric field, since $\left|\textbf{E}_{tot}\right|^{2}$ and $\left|\textbf{B}_{tot}\right|^{2}$ exhibit almost exactly the same barycenter \footnote{The numerically calculated distance between both barycenters is smaller than $1$ nm}. Consequently, we compare the barycenter of $\left|\textbf{E}_{tot}\right|^{2}$ (black cross in \mbox{Fig. \ref{fields}(c)}) and the barycenter of the intensity $S_{z}$ \mbox{(black cross in Fig. \ref{fields}(b))}. Calculations yield a relative shift $\Delta_{S_{z}}$ of approximately $-17$ nm along the $y$-axis. To highlight the relative shifts visually and as a reference value, a dashed white line, representing the $y$-position of the barycenter of $\left|\textbf{E}_{tot}\right|^{2}$ is plotted in \mbox{Fig. \ref{fields}(b-f)}. This relative shift of $S_{z}$ is similar to the results for paraxial beams and a direct consequence of the transverse AM in the focal plane \cite{Aiello2009}. In addition, the transverse AM is indicated by the relative phases of the components of the electric field \mbox{(see Fig. \ref{fields}(d-f))}. No longitudinal spin AM is present, since the phase difference between $E_{x}$ and $E_{y}$, $\Delta \Phi_{x,y}\!=\!\Phi_{x}\!-\!\Phi_{y}$, is an element of $\{-\pi,0,\pi\}$. Furthermore, the absence of a phase vortex for all three components $\Phi_{x}$, $\!\Phi_{y}$, and $\Phi_{z}$ is equivalent to the absence of longitudinal orbital AM. In the barycenter of the focal spot defined via $\left|\textbf{E}_{tot}\right|^{2}$, the electric field vector is rotating around the $x$-axis ($\Delta \Phi_{y,z}\!=\pi/2$), which implies the generation of purely transverse AM \cite{Banzer2013}. \\
Another consequence of the presence of the transverse AM and the observed shift of the barycenter of $S_{z}$ is the deformation of the focal spot. While the electric energy density distribution of the incoming beam $\left|\textbf{E}_{0}\right|^{2}$ is symmetric with respect to the $x$-axis, the symmetry of the focal spot is broken \mbox{(see Fig. \ref{fields}(a) and \ref{fields}(c))}. This asymmetry is caused by the relative shifts of the distributions of the energy density components \mbox{(see Fig. \ref{fields}(d-f))}. Here, the barycenter of $\left|E_{x}\right|^{2}$ is shifted downwards \mbox{($\Delta_{E_{x}}\!\!=\!-33$ nm)}, $\left|E_{y}\right|^{2}$ exhibits almost no shift \mbox{($\Delta_{E_{y}}\!\!=\!3$ nm)}, and $\left|E_{z}\right|^{2}$ is shifted upwards \mbox{($\Delta_{E_{z}}\!\!=\!77$ nm)}. Since the size of the focal spot and the relative shifts have the same order of magnitude, the shape of the focal spot is distinctly and visibly deformed.\\
For the given configuration, the connection of the relative shifts and the deformation is obvious, since in contrast to $\left|E_{y}\right|^{2}$ and $\left|E_{z}\right|^{2}$, $\left|E_{x}\right|^{2}$ has a zero-crossing on the $y$-axis. This focal interference pattern is the justification for using the aforementioned spin-segmented beam. In this experiment, it is therefore sufficient to measure the asymmetric shape of the focal spot to verify the relative shifts of the components of the electric energy density. Beyond that, we reconstruct the focal phase and amplitude distributions for each individual field component, to confirm the gSHEL as origin of the deformation. \\
%%%%%%%%%%%%%%%%%%%%%%%%%%%%%%%%%%%%%%%%%%%%%%%%%%%%%%%%%%%%%%%%%%%%%%%%%%%%%%%%%%%%%%%%%%%%%%%%%%%%%%%%%%%%%%%%%%%%%%%%%%%%%%%%%%
\textit{Experimental approach and results.}\textemdash To measure the deformation of the focal spot and for the reconstruction of the focal field distribution, we use a scanning technique \cite{Banzer2010a} in which a gold nano-sphere (diameter $=90$ nm) sitting on a glass substrate is utilized as a field probe \mbox{(see Fig. \ref{Setup}(a))}. Since the particle is much smaller than the wavelength, it gets excited by the local electric field only, and its polarizability is proportional to the local electric field vector. To suppress the influence of the glass substrate and to guarantee an emission pattern of the excited particle similar to that of a point-like electric dipole, we embed the particle in immersion oil, index matched to the glass substrate. With the above mentioned scheme, the focal spot can be probed by measuring the scattered and transmitted light for each lateral position of the particle relative to the beam axis in the focal plane.\\
The utilized nano-scanning setup is depicted in \mbox{Fig. \ref{Setup}(b)} (see also \cite{Banzer2010a}).
%%%%%%%%%%%%%%%%%%%%%%%%%%%%%%%%%%%%%%%%%%%%%%%%%%%%%%%%%%%%%%%%%%%%%%%%%%%%%%%%%%%%%%%%%%%%%%%%%%%%%%%%%%%%%%%%%%%%%%%%%%%%%%%%%%
%%%%%%%%%%%%%%%%%%%%%%%%%%%%%%%%%%%%%%%%%%%%%%%%%%%%%%%%%%%%%%%%%%%%%%%%%%%%%%%%%%%%%%%%%%%%%%%%%%%%%%%%%%%%%%%%%%%%%%%%%%%%%%%%%%
We start with an incoming $x$-polarized $\text{TEM}_{10}$-mode (see Fig. \ref{Setup}(b)) impinging on a segmented quarter wave plate (SQWP). The two fast axes of the SQWP are perpendicular to each other \cite{Banzer2013}. Therefore, a beam with linear polarization perpendicular to the split-axis is converted into a beam with spatially separated left- and right-handed circular polarization \mbox{(see red arrows in Fig. \ref{Setup}(b))}. By choosing a $\text{TEM}_{10}$-mode as an input beam and overlapping its nodal-line with the split-axis of the SQWP, the diffraction at the split is minimized resulting in a high quality beam profile. The created spin-segmented beam is then strongly focused by an immersion-type microscope objective ($\text{NA}\!\!=\!1.3$). The focal spot is probed by the gold nano-sphere sitting on a glass substrate and embedded in immersion oil. The particle is scanned through the focal plane by a 3D piezo-stage. A second microscope objective ($\text{NA}\!\!=\!1.3$), equivalent to the upper one, is index matched to the glass substrate from below. The objective collects the transmitted light of the tightly focused beam as well as the light scattered forward by the particle. Its back focal plane is imaged with a  CCD-camera. For each position of the particle relative to the beam an image is recorded. The measured data can be used to either directly showcase the deformation of the focal spot or to reconstruct the focal field distribution, including phase and amplitude values of the individual field \mbox{components \cite{Bauer2013}}.\\
First of all, we demonstrate the deformation of the focal spot. For that purpose, we integrate over the intensity pattern for each camera picture. This results in a simple scan image, which assigns one intensity value measured in transmission for each position of the particle. If the wavelength of the beam is adjusted close to the resonance of the particle, the position dependent signal is dominated by the extinction of the beam, resulting in a drop of the signal, when the particle is placed in the focal spot. In \mbox{Fig. \ref{exp:results}(a)} the experimental scan image for the incoming spin-segmented beam is depicted. 
Due to the limited solid angle of the microscope objective, used for collecting the light after interaction with the particle, the scan image cannot be interpreted as an exact description of the total electric energy density in the focal plane. For instance, if the particle is excited by $E_{z}$, only a small amount of the scattered field is collected by the microscope objective. In contrast, an electric dipole excited by $E_{x}$ or $E_{y}$ scatters more dominantly into the collected solid angle. This results in different weightings of the components of the energy density when probing the focal electric field by just scanning the particle through the beam and measuring the intensity of the scattered or extinct field. Still, the measured scan image \mbox{(see Fig. \ref{exp:results}(a))} allows for a comparison with the calculated theoretical distribution of $\left|\textbf{E}_{tot}\right|^{2}$ in \mbox{Fig. \ref{fields}(c)}. However, to verify the presence of transverse AM, the shift of $S_{z}$ and the relative shifts of $\left|E_{x}\right|^{2}$, $\left|E_{y}\right|^{2}$, and $\left|E_{z}\right|^{2}$, the aforementioned reconstruction algorithm is applied \cite{Bauer2013}. The concept is based on the interference between the light scattered off the nanoparticle and the transmitted light beam. Depending on the position of the particle and its local polarization and phase, different interference patterns are measured in the back focal plane of the collecting microscope objective. Via evaluating the back focal plane images for different solid angles, the focal field distribution is reconstructed \footnote{See Supplemental Material and \cite{Bauer2013} for further details of the reconstruction technique.}. The phase and amplitude distributions of the individual electric field components reconstructed from experimental data \mbox{(see Fig. \ref{exp:results}(b-f))} are in very good agreement with the theoretically predicted distributions in \mbox{Fig. \ref{fields}(b-f)}. Slight aberrations, such as the non-planar phase front and the asymmetry with respect to the $y$-axis, are linked to imperfections of the incoming beam and shape deviations of the field probe. For comparison, the relative shifts along the $y$-axis are calculated from the reconstructed distributions to $\Delta_{S_{z}}^{\text{rec}}\!\!=\!-21$ nm, $\Delta_{E_{x}}^{\text{rec}}\!\!=\!-34$ nm, $\Delta_{E_{y}}^{\text{rec}}\!\!=\!4$ nm, $\Delta_{E_{z}}^{\text{rec}}\!\!=\!72$ nm. Also these values are in very good agreement with the theoretically predicted shifts and confirm the appearance of transverse AM and the theory of the gSHEL.\\
%%%%%%%%%%%%%%%%%%%%%%%%%%%%%%%%%%%%%%%%%%%%%%%%%%%%%%%%%%%%%%%%%%%%%%%%%%%%%%%%%%%%%%%%%%%%%%%%%%%%%%%%%%%%%%%%%%%%%%%%%%%%%%%%%%%%%%
\textit{Conclusion.}\textemdash We have discussed the manifestation of the gSHEL in a specially polarized tightly focused vector-beam. The input beam was chosen to be laterally spin-segmented, so that focusing resulted in the generation of transverse AM. As a direct consequence, predicted by the theory of the gSHEL, a deformed focal field distribution was observed. We were able to measure the deformation of the focal spot and verify the relative shifts of the components of the electric energy density as its cause. For that we utilized an experimental nano-probing technique in combination with a reconstruction algorithm. The described and measured appearance of the gSHEL under tight focusing conditions can be relevant for the investigation of polarization dependent effects at the nano-scale, e.g. in nano-plasmonics.
\bibliography{bib}%basename of .bib file
\clearpage
%%%%%%%%%%%%%%%%%%%%%%%%%%%%%%%%%%%%%%%%%%%%%%%%%%%%%%%%%%%%%%%%%%%%%%%%%%%%%%%%%%%%%%%%%%%%%%%%%%%%%%%%%%%%%%%%%%%%%%%%%%%%%%%%%%
%%%%%%%%%%%%%%%%%%%%%%%%%%%%%%%%%%%%%%%%%%%%%%%%%%%%%%%%%%%%%%%%%%%%%%%%%%%%%%%%%%%%%%%%%%%%%%%%%%%%%%%%%%%%%%%%%%%%%%%%%%%%%%%%%%
%%%%%%%%%%%%%%%%%%%%%%%%%%%%%%%%%%%%%%%%%%%%%%%%%%%%%%%%%%%%%%%%%%%%%%%%%%%%%%%%%%%%%%%%%%%%%%%%%%%%%%%%%%%%%%%%%%%%%%%%%%%%%%%%%%
\section{Supplemental information}
In this supplementary text, we discuss the necessary adaptations of the reconstruction technique described \mbox{in \cite{Bauer2013}} to the experiment presented in the main text. The basic systematical difference is the lack of an optical interface, since here, the particle is embedded in a homogenous environment (glass and index matching \mbox{immersion oil}). We start with a short derivation of the reconstruction algorithm.
\subsection{Theoretical considerations}
Let us define a coordinate system $\mathbf{R}=\left(R, \theta , \phi  \right)$, which we relate to the center of the considered beam. In a homogenous medium a source-free vectorial electromagnetic field can be expressed as 
\begin{equation}
\mathbf{E}_{\mathrm{i}}\left(\mathbf{R}\right)=\sum _{n=1}^{\infty} \sum _{m=-n}^n A_{mn}\mathbf{N}^{(1)} _{mn}\left(\mathbf{R}\right) +B_{mn}\mathbf{M}^{(1)} _{mn}\left(\mathbf{R}\right),
\label{eq:E_inc}
\end{equation}
where $A_{mn}$ and $B_{mn}$ are multipole expansion coefficients and $\mathbf{M}^{(1)} _{mn}\left(\mathbf{R}\right)$ and $\mathbf{N}^{(1)} _{mn}\left(\mathbf{R}\right)$ are regular vector spherical harmonics (VSH) \cite{Tsang2000} 
\begin{align}
\mathbf{M}^{(1,3)}_{mn}&= \gamma_{mn}g_n^{(1,3)}\left[\mathbf{e}_{\theta} \frac{\mathrm{i} m}{\sin{\theta}} P_n^m - \mathbf{e}_{\phi}  \frac{\partial}{\partial \theta}P_n^m \right]\mathrm{e}^{\mathrm{i} m\phi} \nonumber \\ 
&= \gamma_{mn}g_n^{(1,3)}\mathbf{X}^{(II)}_{mn}\left(\theta, \phi\right) \text{,}\nonumber \\
\mathbf{N}^{(1,3)}_{mn}&= \gamma_{mn}\left\{\mathbf{e}_Rn(n+1)\frac{g_n^{(1,3)}}{kR}P_n^m + \right. \nonumber \\
& + \left. \frac{\partial\left(R g_n^{(1,3)} \right)}{R\,\partial R} \left[\mathbf{e}_{\theta} \frac{\partial}{\partial \theta}P_n^m + \mathbf{e}_{\phi}\frac{\mathrm{i} m}{\sin{\theta}}P_n^m\right]\right\}\mathrm{e}^{\mathrm{i} m\phi} \nonumber \\ 
&=\gamma_{mn}\left[\mathbf{e}_Rn(n+1)\frac{g_n^{(1,3)}}{kR}P_n^m\mathrm{e}^{\mathrm{i} m\phi} + \right. \nonumber \\ 
&+ \left. \frac{\partial\left(R g_n^{(1,3)} \right)}{R\,\partial R} \mathbf{X}^{(I)}_{mn}\left(\theta, \phi\right) \right] \text{,}
\label{eq:LMPsp}
\end{align}
where $g_n^{(1)}$ represents a spherical Bessel function $j_n(kR)$ and $g_n^{(3)}$ a spherical Hankel \mbox{function $h_n(kR)$} of the first order. The prefactor $\gamma_{mn}=\left[\left(\left(2n+1\right)\left(n-m\right)!\right)/\left(4\pi n\left(n+1\right)\left(n+m\right)!\right)\right]^{1/2}$ and $P_n^m$ are Legendre polynomials with argument $\cos \theta$. The plane wave representation of the first (regular) and third type (irregular) VSH is written as \cite{Tsang2000}
\begin{align}
\mathbf{M}^{(1,3)} _{mn} &= \frac{\left(-\mathrm{i} \right)^n \gamma _{mn}}{2\pi\left(1 + \delta_{(1,3),1} \right)}\int _\Omega \mathrm{d}\Omega \mathrm{e}^{\mathrm{i}\mathbf{k} \cdot \mathbf{R}} \mathbf{X}^{(II)}_{mn}\left(\theta_k, \phi _k \right)\text{,}   \nonumber \\
\mathbf{N}^{(1,3)} _{mn} &= \frac{\left(-\mathrm{i} \right)^{n-1} \gamma _{mn}}{2\pi\left(1 + \delta_{(1,3),1} \right)} \int _\Omega \mathrm{d}\Omega \mathrm{e}^{\mathrm{i}\mathbf{k} \cdot \mathbf{R}} \mathbf{X}^{(I)}_{mn}\left(\theta_k, \phi _k \right)\text{.}
\label{eq:MN_plane}
\end{align}
The Kronecker delta $\delta_{(1,3),1}$ is equal to one for regular VSH and equal to zero for irregular VSH.\\
Now, we consider the position dependent scattering of the spherical gold nano-particle, which is scanned through the light beam in the focal plane. For that purpose, we introduce a new coordinate frame attached to the particle ($R', \theta ', \phi '$), which is related to the original one by $\mathbf{R}=\mathbf{R}'+\mathbf{R_0}$, where $\mathbf{R_0}=\left(R_0, \theta _0, \phi _0 \right)$ is the displacement. The polar ($\theta ' = 0$) and the azimuthal ($\phi '=0$) axes are parallel to the corresponding axes $\theta =0$ and $\phi = 0$. The functions  $\mathbf{M}^{(1)}_{mn}$, $\mathbf{N}^{(1)}_{mn}$ of the original coordinate frame can be expressed as a sum of functions  $\mathbf{M}'^{(1)}_{\mu \nu}$, $\mathbf{N}'^{(1)}_{\mu \nu}$ of the new coordinate frame and the corresponding electric field becomes $\mathbf{E'}_\text{i}=\sum _{\nu=1}^{\infty}\sum _{\mu=-\nu}^{\nu}  A'_{\mu \nu} \mathbf{N}'^{(1)}_{\mu\nu} + B'_{\mu \nu} \mathbf{M}'^{(1)}_{\mu\nu}$. The expansion coefficients $A'_{\mu \nu}$ and $B'_{\mu \nu}$ are found from the VSH addition theorem \cite{Cruzan1962} and can be expressed in the matrix representation \cite{Tsang2000}  
\begin{equation}
\left[ \begin{array}{ll}
	\mathbf{B'} \\
	\mathbf{A'}
	\end{array} \right]= \hat{\mathbf{D}}(\mathbf{R}_0)
	 \left[ \begin{array}{ll}
	\mathbf{B} \\
	\mathbf{A}
	\end{array} \right]\text{,}
\label{eq:add_theor}
\end{equation}
with the displacement operator $\hat{\mathbf{D}}(\mathbf{R}_0)$. The field scattered by the particle can be expressed as a sum of the irregular VSH
\begin{equation}
\mathbf{E'}_\text{s}=\sum _{\nu=1}^{\infty} \sum _{\mu=-\nu}^{\nu}C'_{\mu \nu}\mathbf{N}'^{(3)}_{\mu \nu}\left(\mathbf{R'}\right)  + D'_{\mu\nu}\mathbf{M}'^{(3)}_{\mu\nu}\left(\mathbf{R'}\right), 
\label{eq:E_sca}
\end{equation}
where 
$\mathbf{M}^{(3)} _{\mu\nu}\left(\mathbf{R'}\right)$ and $\mathbf{N}^{(3)} _{\mu\nu}\left(\mathbf{R'}\right)$ are irregular VSH, see (\ref{eq:LMPsp}), and $C_{\mu\nu}$ and $D_{\mu\nu}$ are multipole expansion coefficients of the scattered field. 
Further, for the sake of convenience, we introduce the concept of the T-matrix \cite{Tsang2000}, which relates the vector representations of scattered and incident fields as $\mathbf{E'}_s=\hat{\mathbf{T}}\mathbf{E'}_i$. So, the total electric field $\mathbf{E}_\text{t}=\mathbf{E}_\text{i}+\left(\hat{\mathbf{D}}^{*}\hat{\mathbf{T}}\hat{\mathbf{D}}\right)\mathbf{E}_\text{i}$. \\
The density of the time averaged Poynting vector $\mathbf{P}_\text{t}=1/2 \mathrm{Re} \left[\mathbf{E}_\text{t} \times \mathbf{H}^{*}_\text{t} \right]$ describes the direction of the electromagnetic power flow through a spherical surface element $d\mathbf{\Omega}$. In matrix form and far away from the focal spot it can be represented as
\begin{align}
\mathbf{P}_\text{t} &= \frac{1}{2}\mathrm{Re} \mathbf{E}^{*}_\text{i} \hat{\mathbf{w}}_{t} \mathbf{E}_\text{i}\text{,}\nonumber \\  \label{eq:Poynt3}\hat{\mathbf{w}}_{t}&=\hat{\mathbf{w}}_{b} +\hat{\mathbf{D}}^{*}\hat{\mathbf{T}}^{*}\hat{\mathbf{w}}_{s}\hat{\mathbf{T}}\hat{\mathbf{D}}+\hat{\mathbf{D}}^{*}\hat{\mathbf{T}}^{*}\hat{\mathbf{w}}_{e}+\hat{\mathbf{w}}_{e}\hat{\mathbf{T}}\hat{\mathbf{D}}\text{.}
\end{align}
Here $\hat{\mathbf{w}}_{i}$, $\hat{\mathbf{w}}_{s}$, $\hat{\mathbf{w}}_{e}$ are azimuthally ($\theta$) and polarly ($\phi$) dependent operators
\begin{widetext}
\begin{align}
 \hat{\mathbf{w}}_{i}= \frac{\mathrm{i}}{2k^2}\sqrt{\frac{\epsilon}{\mu}}\left[ 
 \begin{array}{ll}
	\mathrm{i}\sin\left(\frac{\nu_1-\nu_2}{2}\pi\right)\mathbf{X}^{(II)}_{\mu_1\nu_1}\cdot \mathbf{X}^{*(II)}_{\mu_2\nu_2} &\cos\left(\frac{\nu_1-\nu_2}{2}\pi \right) \mathbf{X}^{(I)}_{\mu_1\nu_1}\cdot \mathbf{X}^{*(II)}_{\mu_2\nu_2}  \\
	-\cos\left(\frac{\nu_1-\nu_2}{2}\pi \right)\mathbf{X}^{(II)}_{\mu_1\nu_1}\cdot \mathbf{X}^{*(I)}_{\mu_2\nu_2}   & \mathrm{i}\sin\left(\frac{\nu_1-\nu_2}{2}\pi\right)\mathbf{X}^{(I)}_{\mu_1\nu_1}\cdot \mathbf{X}^{*(I)}_{\mu_2\nu_2}  
	\end{array} \right],\nonumber \\	
\hat{\mathbf{w}}_{s}= \frac{\mathrm{i}}{k^2}\sqrt{\frac{\epsilon}{\mu}}\left[ 
\begin{array}{ll}
	\mathrm{i}^{\nu_2-\nu_1+1}\mathbf{X}^{(II)}_{\mu_1\nu_1}\cdot \mathbf{X}^{*(II)}_{\mu_2\nu_2} &\mathrm{i}^{\nu_2-\nu_1} \mathbf{X}^{(I)}_{\mu_1\nu_1}\cdot \mathbf{X}^{*(II)}_{\mu_2\nu_2}  \\
	-\mathrm{i}^{\nu_2-\nu_1}\mathbf{X}^{(II)}_{\mu_1\nu_1}\cdot \mathbf{X}^{*(I)}_{\mu_2\nu_2}   & \mathrm{i}^{\nu_2-\nu_1+1}\mathbf{X}^{(I)}_{\mu_1\nu_1}\cdot \mathbf{X}^{*(I)}_{\mu_2\nu_2}  
	\end{array}
	 \right] , \quad
 \hat{\mathbf{w}}_{e}= \frac{1}{2}\hat{\mathbf{w}}_{s}.
\label{eq:w_ise}
\end{align}
\end{widetext}
The energy transmitted into the solid angle $\Omega=\left[\theta \in \left(\theta_1, \theta_2 \right), \phi \in \left(\phi_1, \phi_2 \right) \right]$ is
\begin{align}
T \left(\phi_1, \phi_2, \theta_1, \theta_2 \right) = \int _{ \Omega} d\Omega  \mathbf{P}_\text{t}\left(\theta, \phi \right) 
\label{eq:TR_air}
\end{align}
or in a matrix form
\begin{align}
T \left(\phi_1, \phi_2, \theta_1, \theta_2 \right) = \mathrm{Re} \mathbf{E}^{*}_\text{i} \hat{\mathbf{W}}_{t} \mathbf{E}_\text{i} \text{.}
\label{eq:TR_air2}
\end{align}
Here, $\hat{\mathbf{W}}_{t}$ denotes the operator $\hat{\mathbf{w}}_{t}$, whose elements were integrated over a region $\Omega$:
\begin{align}
\hat{\mathbf{W}}_{t} = \int _{\Omega} d\Omega\hat{\mathbf{w}}_{t} \text{.}
\label{eq:TR_air25}
\end{align}
\subsection{Experimental implementation}
\begin{figure*}[h!]
\includegraphics[width=1\textwidth]{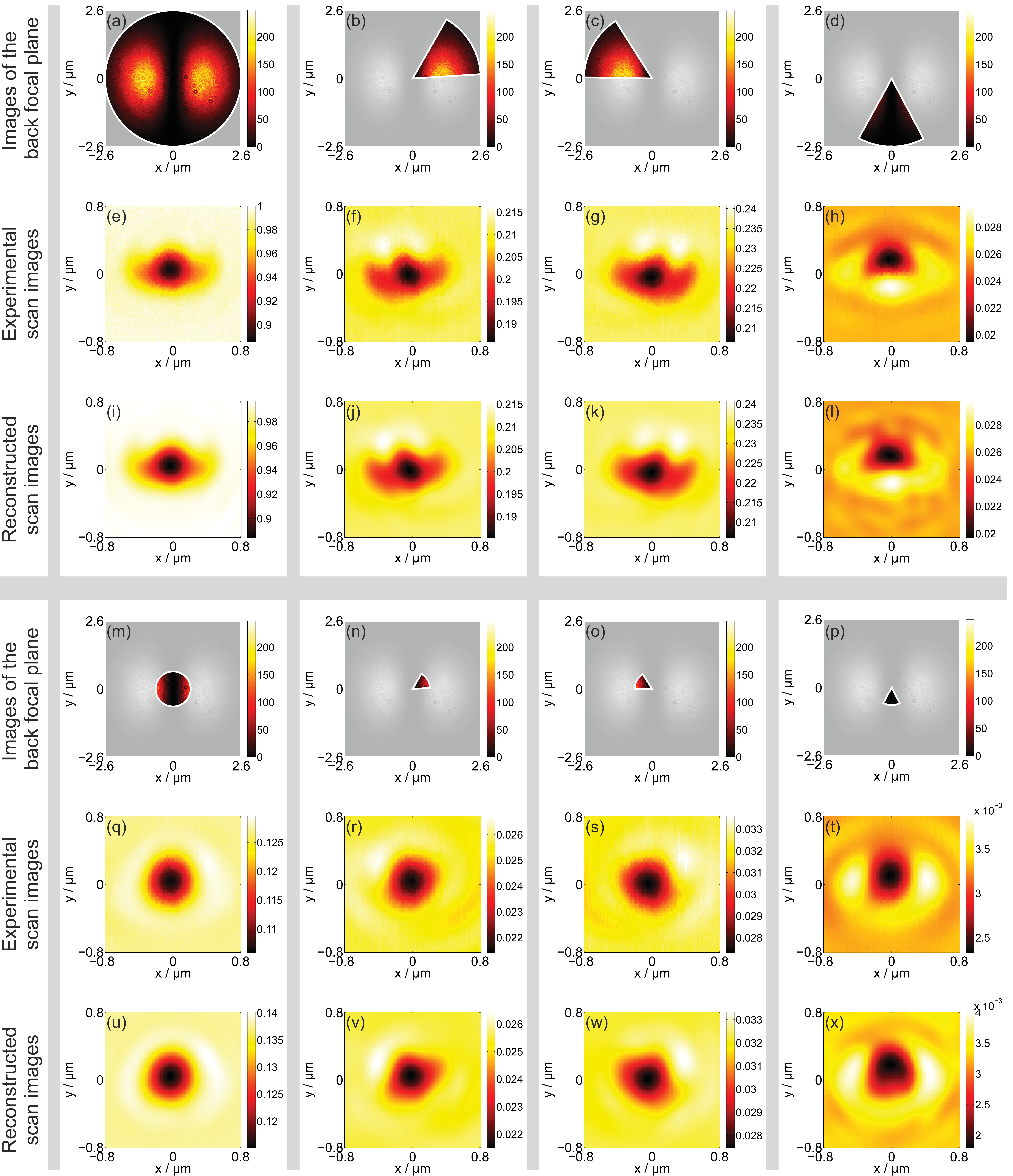}
\caption{\label{Rec} (a-d) and (m-p) show images of the back focal plane. Different integration areas are highlighted with a white frame. (e-h) and (q-t) represent the associated experimental scanning images. Each pixel of a scanning image represents the measured power value within the integration area for the corresponding particle position. (i-l) and (u-x) are the fitted distributions.}
\end{figure*}

Equation (\ref{eq:TR_air2}) is the theoretical foundation of the reconstruction technique. Since the parameters of the gold particle are known for the given wavelength $\lambda=590$ nm (diameter $=90$ nm, $\epsilon=-5.67+ 1.14\mathrm{i}$), we can calculate the matrix elements of $\hat{\mathbf{w}}_{t}$. From the experimental point of view, it is possible to access $T(\Omega)$ via imaging the back focal plane of the microscope objective in transmission with a CCD-camera (see main text). $\Omega$ is limited by the solid angle of the microscope objective. We measure $T(\Omega)$ for eight different volume angles simply by integrating the measured intensity over the corresponding angular region \mbox{(see Fig. \ref{Rec}(a-d) and \ref{Rec}(m-p))}. For one particle position relative to the beam, this results in a number of eight equations. However, the number of equations can be increased drastically via repeating the same measurement for different particle positions $R_{0}$. The actual reconstruction is performed as follows:\\
The particle is scanned through the focal plane of the beam. For each particle position an image of the back focal plane is recorded. The step size between each position is chosen to be $16$ nm, whereby the size of the scan field is $1.6\times1.6\,\mu m^{2}$. A total number of $10201$ images are recorded. For each position/image, we integrate over the eight volume angles, resulting in eight power values. Those values can be reassembled to eight scanning images (see Fig. \ref{Rec}(e-h) and \ref{Rec}(q-t)). In a last step, we fit multipole expansion coefficients to the experimental data. Here, we consider multipoles up to the order of eight. The resulting fitted distributions can be seen in \mbox{Fig. \ref{Rec}(i-l) and \ref{Rec}(u-x)}. The good overlap between the experimental scan images and the fitted distributions shows, that the fit has been successful.\\
To compare the reconstructed beam with the calculated distributions of $\left|\textbf{E}_{tot}\right|^{2}$, $\left|E_{x}\right|^{2}$, $\left|E_{x}\right|^{2}$ and $\left|E_{z}\right|^{2}$ (see main text Fig. 2), we insert the fitted multipole expansion coefficients in equation (\ref{eq:E_inc}). The \mbox{$z$-component} of the Poynting vector $S_{z}$ is proportional to $\mathrm{Re}\left[\textbf{E}\times\left(\mathrm{i}\nabla\times\textbf{E}\right)^{*}\right]_{z}$. The results are plotted in Fig. 4 in the main text.
%\begin{thebibliography}{99}
%\bibitem{Bauer2013} T. Bauer, S. Orlov, U. Peschel, P. Banzer, G. Leuchs, arXiv:1304.4444v1
%\bibitem{LTsang} L. Tsang, J. A. Kong, K.-H. Ding "Scattering of electromagnetic waves", John Wiley, New York.
%\bibitem{ORCruz62} O. R. Cruzan, Q. Appl. Math., 20 (1962) 33.
%\end{thebibliography} 
\end{document}